\documentclass[reprint,superscriptaddress, 
aps,preprintnumbers]{revtex4-1}
\usepackage{bm,bbm}
\usepackage{natbib,textcase}
\usepackage{breakurl}
\usepackage{booktabs}
\usepackage{multirow}
\usepackage{mathtools, amsfonts, amsthm, latexsym, amssymb,dsfont}
\usepackage{braket}
\usepackage[T1]{fontenc}
\usepackage{microtype}
\usepackage[hyperindex,breaklinks]{hyperref}
\usepackage{graphicx}% Include figure files
\usepackage{caption}
\usepackage{subcaption}
\usepackage{xcolor}
\usepackage{array}
\usepackage{enumitem}

\hypersetup{
    breaklinks=true,   % splits links across lines
    colorlinks=true,
    linkcolor=blue,
    filecolor=magenta,      
    urlcolor=cyan,
    pdfpagemode=FullScreen,
    }

\newtheoremstyle{break}
  {\topsep}{\topsep}%
  {\upshape}{}%
  {\bfseries}{}%
  {\newline}{}%
\theoremstyle{break}

\def\Tr{{\rm Tr}}

\def\i{{\rm i}}
\def\CB{{\cal B}}
\def\CC{{\cal C}}

\def\CF{{\cal F}}

\def\CH{{\cal H}}

\def\CN{{\cal N}}

\def\BC{\mathbb{C}}
\def\BF{\mathbb{F}}

\def\BR{\mathbb{R}}

\def\BZ{\mathbb{Z}}
\def\Ba{\mathbf{a}}
\def\Bb{\mathbf{b}}

\def\Balpha{\boldsymbol{\alpha}}
\def\Bbeta{\boldsymbol{\beta}}
\def\BOmega{\boldsymbol{{\Omega}}}
\def\Bzero{\boldsymbol{{0}}}
\def\Bdelta{\boldsymbol{{\delta}}}
\def\BDelta{\boldsymbol{{\Delta}}}

\begin{document}
\title{Supersymmetric conformal field theories from quantum stabilizer codes}

\author{Kohki Kawabata}
\affiliation{Department of Physics, Faculty of Science,
The University of Tokyo,\\
Bunkyo-Ku, Tokyo 113-0033, Japan}
\affiliation{
Department of Physics, Osaka University,\\
Machikaneyama-Cho 1-1, Toyonaka 560-0043, Japan
}

\author{Tatsuma Nishioka}
\affiliation{
Department of Physics, Osaka University,\\
Machikaneyama-Cho 1-1, Toyonaka 560-0043, Japan
}

\author{Takuya Okuda}
\affiliation{
Graduate School of Arts and Sciences, The University of Tokyo, \\
Komaba, Meguro-ku, Tokyo 153-8902, Japan
}

\date{\today}

\preprint{OU-HET-1195, UT-Komaba/23-9}

\begin{abstract}
We construct fermionic conformal field theories (CFTs) whose spectra are characterized by quantum stabilizer codes.
We exploit our construction to search for fermionic CFTs with supersymmetry by focusing on quantum stabilizer codes of the Calderbank-Shor-Steane type, and derive simple criteria for the theories to be supersymmetric.
We provide several examples of fermionic CFTs that meet the criteria, and find quantum codes that realize
$\CN =4$ supersymmetry.
Our work constitutes a new application of quantum codes and paves the way for the methodical search for supersymmetric CFTs.
\end{abstract}

\maketitle

\section{Introduction}
The past two decades have seen a growing interest in quantum information theory as a foundation for quantum computing and broad applications to various branches of theoretical physics.
In particular, quantum error correction (QEC) is the key to the experimental realization of fault-tolerant quantum computers robust against quantum noises such as decoherence \cite{Shor:1995hbe,Steane:1996va,Knill:1996ny,Shor:1996qc,nielsen_chuang_2010}.
QEC codes are the theoretical framework that protects quantum states (code subspaces) from errors by embedding them into larger Hilbert spaces.
In condensed matter physics, a large class of QEC codes is constructed to describe topological phases of matter such as toric codes~\cite{Kitaev:1997wr,Bullock:2006bv,Levin:2004mi} and fracton models~\cite{Chamon:2004lew,Haah:2011drr,Vijay:2015mka,Vijay:2016phm} as code subspaces.
On the other hand, holographic codes \cite{Almheiri:2014lwa,Pastawski:2015qua,Hayden:2016cfa,Pastawski:2016qrs} have been investigated in high energy theory to understand holographic duality between quantum gravity and quantum field theory in one lower dimensions \cite{tHooft:1993dmi,Susskind:1994vu,Maldacena:1997re}. 

QEC codes have been exploited to construct  a discrete set of two-dimensional conformal field theories (CFTs), called Narain code CFTs \cite{Dymarsky:2020qom}.
This generalizes the construction of chiral CFTs from classical codes \footnote{Classical codes can be used to construct Narain CFTs \cite{Dymarsky:2021xfc,Yahagi:2022idq}, but in this letter we call CFTs built out of quantum codes Narain code CFTs.}, which has been known for a long time \cite{frenkel1989vertex,Dolan:1994st}.
Narain code CFTs are bosonic CFTs 
whose spectra are characterized by Lorentzian lattices associated with quantum stabilizer codes.
Narain code CFTs find their applications in the modular bootstrap program \cite{Dymarsky:2020bps,Henriksson:2022dnu,Dymarsky:2022kwb}, the search for CFTs with large spectral gaps \cite{Furuta:2022ykh,Angelinos:2022umf}, and holographic duality \cite{Dymarsky:2020pzc} based on ensemble average \cite{Maloney:2020nni,Afkhami-Jeddi:2020ezh}.
More recently, Narain code CFTs have been generalized from qubit (binary) to qudit (non-binary) stabilizer codes \cite{Kawabata:2022jxt,Alam:2023qac}, and a family of code CFTs is constructed from quantum Calderbank-Shor-Steane (CSS) codes~\cite{Calderbank:1995dw,Steane:1996va}.
See also \cite{Buican:2021uyp,Henriksson:2022dml,Kawabata:2023nlt,Furuta:2023xwl} for other developments.

In this letter, we expand on the prescriptions of \cite{Dymarsky:2020qom,Kawabata:2022jxt} and construct fermionic CFTs from quantum stabilizer codes.
Our strategy is to use the modern description~\cite{Tachikawalec,Karch:2019lnn,Hsieh:2020uwb} of fermionization \cite{Ginsparg:1988ui}, which turns a bosonic theory with a $\BZ_2$ symmetry into a fermionic theory.
We will describe the fermionization of a Narain code CFT in terms of the modification of the Lorentzian lattice underlying the CFT.
Our formulation 
makes manifest the relationship between the sectors of fermionic CFTs and the modified lattices.

Furthermore, we leverage our construction to search for supersymmetric CFTs, i.e., fermionic CFTs with supersymmetry.
The emergence of supersymmetry has attracted much attention
in high energy theory \cite{Parisi:1979ka,Dixon:1988qd,Harvey:2020jvu,Bae:2020pvv,Bae:2021jkc,Bae:2021lvk,Kaviraj:2021qii} and even in condensed matter physics \cite{Lee:2006if,Grover:2012bm,Grover:2013rc,Rahmani:2015qpa,Hsieh:2016emq,Jian:2016zll,Ma:2021dua}.
In the chiral case, there are notable examples of supersymmetric CFTs built out of classical codes \cite{Dixon:1988qd,Gaiotto:2018ypj,Kawabata:2023nlt}, but no analog has been known in the non-chiral case.
In this letter, we examine when the fermionic code CFTs constructed from quantum CSS codes are supersymmetric.
We derive simple criteria for supersymmetry that can be tested for a given CSS code.
By applying the criteria to the codes of small lengths, we find two codes of length six that yield the same fermionic CFT satisfying the conditions for supersymmetry, and moreover prove that the resulting theory is equivalent to a description of the K3 sigma model with $\CN = 4$ supersymmetry \cite{Gaberdiel:2013psa}.
We find more candidates of supersymmetric code CFTs and leave it as an open problem to establish the actual existence of supersymmetry.
Our results signify the ubiquity of QEC codes as a universal structure in theoretical physics.

\section{Narain code CFTs}

We start with reviewing the construction of bosonic CFTs from quantum stabilizer codes \cite{Dymarsky:2020qom,Kawabata:2022jxt}.
Let $\BF_p = \BZ/p\,\BZ$ be a finite field for an odd prime $p$ and $\{\, \ket{x}\,\big|\, x\in \BF_p\, \}$ an orthonormal basis for the Hilbert space $\BC^p$. 
The Pauli group acting on $\BC^p$ is generated by the operators $X$ and $Z$ 
defined by $X\,\ket{x} = \ket{x+1},  Z\,\ket{x} = \omega^x\,\ket{x}$,
where $\omega = \exp(2\pi \i/p)$ and $x$ is defined modulo $p$ \cite{Gottesman:1997zz}.
$X$ and $Z$ are generalizations of the Pauli matrices acting on a qubit system.
For an $n$-qudit system, we define 
$
    g(\Ba, \Bb) \equiv
    X^{a_1} Z^{b_1}\otimes \cdots \otimes X^{a_n} Z^{b_n}
$, where $\Ba = (a_1, \cdots, a_n), \Bb = (b_1, \cdots, b_n)
\in \BF_p^n$.
While a pair of two 
operators $g(\Ba, \Bb)$ and $g(\Ba', \Bb')$ do not commute in general, they commute if $\Ba\cdot \Bb' - \Ba'\cdot \Bb = 0$, where $\Ba\cdot\Bb = \sum_{i=1}^n a_i\,b_i$.
An $[[n, k]]$ quantum stabilizer code $V_{\mathsf{S}}$ is defined as a $p^k$-dimensional subspace of $(\BC^p)^{n}$ fixed by the stabilizer group $\mathsf{S} = \langle g_1, \cdots, g_{n-k}\rangle$ generated by a commuting set of 
operators $g_i = g(\Ba^{(i)}, \Bb^{(i)})~(i=1, \cdots, n-k)$~\cite{Gottesman:1996rt,Gottesman:1998se,knill1996non,knill1996group,rains1999nonbinary}.

There is an intriguing relation between quantum stabilizer codes and classical codes \cite{Calderbank:1996aj,ashikhmin2001nonbinary}.
Consider a classical code $\CC$ specified by the stabilizer group $\mathsf{S}$:
\begin{align}
    \CC = \left\{\, c=(\Ba,\Bb)\in \BF_p^{2n}\;\middle|\; g(\Ba,\Bb)\in \mathsf{S}\,\right\}\ .
\end{align}
The classical code $\CC$ admits the Lorentzian inner product taking values in $\BF_p$: $c\odot c' = c\,\eta\,c'\,^T$ ($c, c'\in \CC$) where
\begin{align}
\label{eq:metric}
    \eta 
        =
        \left[
        \begin{array}{cc}
    	0 & I_{n} \\
            I_n & 0
        \end{array}
        \right] \ .
\end{align}
The associated dual code $\CC^\perp$ consists of elements $c'\in \BF_p^{2n}$ satisfying $c'\odot c=0$ mod $p$ for any $c\in \CC$.
$\CC$ is called self-orthogonal if $\CC\subset \CC^\perp$ and self-dual if $\CC = \CC^\perp$.
For a qudit stabilizer code, $\CC$ is not necessarily self-dual.
However, for self-dual codes $\CC$, the Construction A lattice \cite{conway2013sphere}
\begin{align}
    \Lambda(\CC) \equiv \left\{ \frac{c + p\,m}{\sqrt{p}}\, \bigg|\, c\in \CC, ~ m\in \BZ^{2n} \right\}
\end{align}
is even self-dual with respect to the Lorentzian inner product $\lambda\odot\lambda'\equiv\lambda\,\eta\,\lambda'\,^T$: $\lambda\odot\lambda\in2\BZ$ for any $\lambda\in \Lambda(\CC)$ and there is one lattice point per unit volume.

The Construction A lattice~$\Lambda(\CC)$ is related to the momentum lattice~$\widetilde{\Lambda}(\CC)$ of the Narain CFT \cite{Narain:1985jj,Narain:1986am}
via $(p_L,p_R) = (\lambda_1 + \lambda_2, \lambda_1 - \lambda_2)/\sqrt{2}\in \widetilde{\Lambda}(\CC)$, where $(\lambda_1,\lambda_2)\in \Lambda(\CC)$.
The vertex operators in a Narain code CFT are constructed from 
$(p_L,p_R)\in \widetilde{\Lambda}(\CC)$ as
$
    V_{p_L,p_R}(z,\bar{z}) =\, \exp \left( \i\, p_L\cdot X_L(z) + \i\, p_R\cdot X_R(\bar{z})\right)
$,
where $X(z,\bar{z}) =X_L(z) + X_R(\bar{z})$ is an $n$-dimensional free boson.
The resulting theory is a bosonic CFT of central charge $n$, whose consistency is guaranteed by the evenness and the self-duality of the Construction A lattice.
Using the state-operator isomorphism, the vertex operators correspond to the momentum states $\ket{p_L,p_R}$.
They are eigenstates of the Virasoro generators $L_0$ and $\bar L_0$ with the eigenvalues (conformal weights) $h = p_L^2/2$ and $\bar h = p_R^2/2$, respectively.
Taking account of the excitation by bosonic oscillators, we obtain the whole Hilbert space $\CH(\CC)$ of the Narain code CFT.

We measure the spectrum of the Narain code CFT by the torus partition function $Z_{\CC}$ defined by
\begin{align}
    \begin{aligned}
    \frac{Z_{\CC}(\tau, \bar\tau)}{(q\bar q)^{- \frac{n}{24}}}
        &= 
        \Tr_{\CH(\CC)}\left[ q^{L_0}\,\bar q^{\bar L_0}\right] 
        =\!\!\!\!\!\!
         \sum_{(p_L, p_R) \in \tilde\Lambda(\CC)} q^{\frac{p_L^2}{2}}\,\bar q^{\frac{p_R^2}{2}},
    \end{aligned}
\end{align}
where $\tau=\tau_1+\i\tau_2$ is the modulus of the torus and $q=\exp(2\pi\i\tau)$.
Let us define the complete weight enumerator~\cite{macwilliams1977theory} of the self-dual code $\CC\subset\BF_p^{2n}$ as
\begin{align}
    W_{\CC}\left(\{x_{ab}\}\right) = \sum_{c\,\in\,\CC} \;\prod_{(a,b)\,\in\,\BF_p\times\BF_p} x_{a b}^{\mathrm{wt}_{ab}(c)}\,,
\end{align}
where $\mathrm{wt}_{ab}(c) = \left|\left\{i\,|\, (c_i,c_{i+n}) = (a,b)\right\}\right|$.
The torus partition function of the CFT can be written using the complete weight enumerator:
\begin{align}
    Z_{\CC}(\tau,\bar{\tau}) = \frac{1}{|\eta(\tau)|^{2n}} \,W_{\CC}(\{\psi_{ab}^+\})
\end{align}
with $\eta(\tau)$ the Dedekind eta function and 
$
    \psi_{ab}^+
        = 
            \Theta\genfrac{[}{]}{0.0pt}{}{\Balpha}{\Bzero}
            \left(\Bzero\,\middle|\,\BOmega\right)
$,
where 
\begin{align}
    \Balpha 
            =
            \left( \frac{a}{p}, \frac{b}{p}\right), \quad
        \BOmega
            =
            p\begin{bmatrix}
                    \i\,\tau_2 & \tau_1 \\
                    \tau_1 & \i\, \tau_2
                \end{bmatrix},
\end{align}
and
$\Theta$ is the Siegel theta function of genus-two defined by
\begin{align}
    \Theta\genfrac{[}{]}{0.0pt}{}{\Balpha}{\Bbeta}%
\left(\mathbf{z}\,\middle|\,\BOmega\right)
        =\!
        \sum_{\mathbf{n}\in%
        \BZ^{2}}e^{2\pi \i\left[\frac{(\mathbf{n}+\Balpha)%
        \boldsymbol{{\Omega}}(\mathbf{n}+\Balpha)^T}{2}+(\mathbf{n}+%
        \Balpha)(\mathbf{z}+\Bbeta)^T\right]}.
\end{align}

\section{Fermionization and lattice modification}

Consider a bosonic CFT $\CB$ with a global non-anomalous $\BZ_2$ symmetry $\sigma$.
The Hilbert space $\CH$ can be decomposed to the even and odd subsectors under the $\BZ_2$ as $\CH = \CH^+\oplus \CH^-$, where $\CH^\pm \equiv \{ \psi \in \CH\, |\, \sigma\,\psi = \pm \psi\}$.
$\CH$ will be called the untwisted sector.
To define the twisted sector, let us place the theory on a cylinder.
In the $\sigma$-twisted Hilbert space $\CH_\sigma$, a field $\phi$ obeys the twisted boundary condition $\CH_\sigma: \phi(x + 2\pi) = \sigma\, \phi(x)$,
along the circle direction parametrized by $x$.
The twisted sector also decomposes to the $\BZ_2$ even and odd subsectors: $\CH_\sigma = \CH_\sigma^+\oplus \CH_\sigma^-$.
One can define the partition function $S$ of the $\BZ_2$-even sector $\CH^+$ by $S = (q\bar q)^{ - \frac{n}{24}}\Tr_{\CH^+}\left[ q^{L_0}\,\bar q^{\bar L_0}\right]$.
The three partition functions $T,U$, and $V$ can be defined similarly for the other sectors $\CH^-, \CH_\sigma^+$ and $\CH_\sigma^-$, respectively.
Then the partition function of (the untwisted sector of) $\CB$ can be written as $Z_\CB = S + T$.

A fermionic theory $\CF$ can be constructed from the bosonic theory $\CB$ by coupling the latter to
a spin topological quantum field theory with a $\BZ_2$ symmetry and gauging the diagonal $\BZ_2$ symmetry.
This is the modern description of fermionization \cite{Karch:2019lnn,Hsieh:2020uwb,Tachikawalec}.
The Hilbert space of the resulting fermionic theory on a circle has Neveu-Schwarz (NS) and Ramond (R) sectors~\footnote{%
A fermionic theory is one that depends on the spin structure of spacetime.  On a spatial circle, fermionic operators are anti-periodic in the NS sector and periodic in the R sector.
}. 
Each of the NS and R sectors decomposes to the $\BZ_2$-even and odd subsectors corresponding to the fermion parity.
Thus there are in total four subsectors $\text{NS} +$, $\text{NS} -$, $\text{R} +$, and $\text{R} -$, whose partition functions are given by $Z_\CF^{\text{NS} +}= S , ~ Z_\CF^{\text{NS} -} = V, ~ Z_\CF^{\text{R} +} = T,~ Z_\CF^{\text{R} -}= U$, respectively.

We now turn to a Narain code CFT for a quantum stabilizer code and its fermionization.
The spectrum of the code CFT is uniquely characterized by the underlying lattice $\Lambda(\CC)$, hence we will recast the fermionization procedure in terms of lattices.
First, let us consider a vector of length $2n$, $\chi = \sqrt{p}\,\textbf{1}_{2n}$, where we introduced the notation $\mathbf{1}_{2n} = (1,1,\cdots, 1)$.
This vector belongs to $\Lambda(\CC)$ but its half $\delta \equiv \chi/2$ does not: $\delta \notin \Lambda(\CC)$.
Then, we decompose $\Lambda(\CC)$ 
as $\Lambda(\CC) = \Lambda_0 \cup \Lambda_1$,
where 
\begin{align}
    \label{eq:lambda0lambda1}
    \Lambda_i = \left\{\lambda\in\Lambda(\CC)\,|\,\chi\odot \lambda = i \;\;\mathrm{mod}\;2\right\}\quad (i=0,1)\ .
\end{align}
We define the $\BZ_2$ symmetry by letting $\Lambda_0$ and $\Lambda_1$ correspond to the Hilbert spaces $\CH^+$ and $\CH^-$, respectively.
Note that $\Lambda_0$ is a sublattice of $\Lambda(\CC)$, while $\Lambda_1$ is not a lattice by itself.
This structure precisely matches the fact that the operators in $\CH^+$ are closed under the operator product expansion (OPE) while those in $\CH^-$ are not.
Let us move to the twisted sector.
We assume $n\in 2\BZ$ to ensure that the $\BZ_2$ symmetry defined by $\chi$ is non-anomalous \footnote{This statement can be shown using the method of \cite{Lin:2019kpn}}.
We then introduce two additional sets by 
\begin{align}\label{eq:lambda2}
    (\Lambda_2, \Lambda_3) &= 
    \begin{dcases}
    (\Lambda_1 + \delta, \Lambda_0 + \delta) & \quad (n\in4\BZ)\,,\\
    (\Lambda_0 + \delta, \Lambda_1 + \delta) & \quad (n\in 4\BZ+2)\,.
    \end{dcases}
\end{align}
One can check that $\Lambda_\text{NS} \equiv \Lambda_0 \cup \Lambda_2$ is a self-dual lattice that is not even (i.e., odd self-dual) with respect to the Lorentzian inner product $\odot$.
The oddness and the self-duality of $\Lambda_\text{NS}$ imply that the spectrum of the associated CFT includes both bosonic and fermionic operators that are closed under OPE.
Hence we can identify $\Lambda_0$ and $\Lambda_2$ with the $\BZ_2$ even and odd subsectors of the NS sector ($\text{NS}+$ and $\text{NS}-$) in the fermionized theory, respectively.
Similarly, $\Lambda_1$ and $\Lambda_3$ correspond to the $\BZ_2$ even and odd subsectors of the R sector.
Table~\ref{tab:fermionization} summarizes the relations between the sectors of code CFTs and~$\Lambda_i$ ($i=0,1,2,3$).

\begin{table}[th]
    \centering
    \begin{subtable}[t]{0.22\textwidth}
    \centering
        \begin{ruledtabular}
            \begin{tabular}{ccc}
                $\CB$   & untwisted & twisted  \\\hline \\[-0.25cm] 
                even    & $S/\Lambda_0$       & $U/\Lambda_3$ \\[0.1cm]
                odd     & $T/\Lambda_1$       & $V/\Lambda_2$
            \end{tabular}
        \end{ruledtabular}
    \caption{Bosonic code CFT}
    \end{subtable}
    \hspace{\fill}
    \begin{subtable}[t]{0.22\textwidth}
        \centering
        \begin{ruledtabular}
            \begin{tabular}{ccc}
                $\CF$   & \quad NS \quad &  \quad R  \quad  \\ \hline \\[-0.25cm] 
                even    & \quad $S/\Lambda_0$ \quad      & \quad $T/\Lambda_1$ \quad \\[0.1cm]
                odd     & \quad $V/\Lambda_2$ \quad      & \quad $U/\Lambda_3$ \quad
            \end{tabular}
        \end{ruledtabular}
    \caption{Fermionic code CFT}
    \end{subtable}
    \caption{The sectors of bosonic and fermionic code CFTs and their relations to $\Lambda_i~(i=0,1,2,3)$.}
    \label{tab:fermionization}
\end{table}

The partition function of each sector of the code CFTs can be calculated from $\Lambda_i~(i=0,1,2,3)$ shown in Table \ref{tab:fermionization}, thus can be represented by the weight enumerator of the associated code $\CC$ \cite{Dymarsky:2020qom,Kawabata:2022jxt}.
By reading off the spectra of the sectors from the norms of the lattice vectors, we find 
\begin{align}
    S \pm T = \frac{ W_\CC (\{\psi_{ab}^\pm\})}{|\eta(\tau)|^{2n}} , \quad 
    U \pm V = \frac{W_\CC (\{\tilde{\psi}_{ab}^\pm\})}{|\eta(\tau)|^{2n}} \ ,
\end{align}
where $\psi_{ab}^-
        =  
            \Theta\genfrac{[}{]}{0.0pt}{}{\Balpha}{\Bzero}
            \left(p\,\Bdelta\,\middle|\,\BOmega\right), 
    \tilde\psi_{ab}^+
        = 
            \Theta\genfrac{[}{]}{0.0pt}{}{\Balpha + \Bdelta}{\Bzero}
            \left(\Bzero\,\middle|\,\BOmega\right),$
    and
    $
    \tilde\psi_{ab}^-
        =  
            \Theta\genfrac{[}{]}{0.0pt}{}{\Balpha + \Bdelta}{\Bzero}
            \left(\Bzero\,\middle|\,\BOmega+\BDelta\right)$
with the parameters 
\begin{align}
        \Bdelta
            =
            \left(\frac{1}{2}, \frac{1}{2}\right),\quad
        \BDelta
            =
            \begin{bmatrix}
                    0 & p \\
                    p & 0
                \end{bmatrix}.
\end{align}

\section{Searching for supersymmetric theories}
Having yielded a general construction
of fermionic CFTs from quantum stabilizer codes, we will exploit our construction to search for supersymmetric CFTs.
Fermionic CFTs with supersymmetry must meet the following conditions (see e.g.,\,\cite{Bae:2021lvk}):
\begin{enumerate}[label=(\roman*)]
    \item The NS sector contains a Virasoro primary with the conformal weight $\left(3/2,0\right)$, $\left(0,3/2\right)$.
    \item The R sector satisfies the positive energy condition $h$, $\bar{h}\geq \frac{n}{24}$.
    \item The Ramond-Ramond (RR) partition function $Z_{\mathrm{RR}}\equiv Z_\CF^{\textrm{R}+} - Z_\CF^{\textrm{R}-}$ is constant.
\end{enumerate}
The first condition is necessary for the existence of operators that generate supersymmetry with conformal weight $3/2$.
The second condition is imposed by the unitarity of the theory.
The third one follows from the cancellation of the contributions from bosonic and fermionic states other than the vacuum in supersymmetric theories.
Recently, the first two conditions have been proven to imply the third in \cite{Bae:2021jkc}, thus fermionic CFTs satisfying the conditions $(\textrm{i}), (\textrm{ii})$ are conjectured to be SCFTs \cite{Bae:2020pvv,Bae:2021lvk,Bae:2021jkc}.

In what follows, we focus on the fermionic CFT constructed from the CSS code corresponding to $\CC = C\times C$, where $C$ is a classical self-dual code with respect to the Euclidean inner product.
The Lorentzian even self-dual lattice for the bosonic CFT takes the form:
\begin{align}\label{eq:Lambda-CSS}
    \Lambda(\CC) = \left\{\left(\frac{c_1 + p\,m_1}{\sqrt{p}},\frac{c_2 + p\,m_2}{\sqrt{p}}\right)\in\BR^{2n}\,\right\},
\end{align}
where $c_1\,,c_2\in C,\;\,m_1, m_2\in\BZ^n$.
Since the resulting non-chiral CFT is left-right symmetric, it is enough to examine the conditions $(\textrm{i}), (\textrm{ii})$ for the left-moving sector.

Let us first examine the condition $(\textrm{i})$.
The spectrum of the primary operators in the NS sector is determined by $\Lambda_\text{NS} = \Lambda_0\cup \Lambda_2$ (see Table \ref{tab:fermionization}).
If there exist primary operators of weight $(h, \bar h) = (3/2, 0)$ in the NS sector, they must be built out of the lattice vectors in $\Lambda_2$ as $h-\bar h = (p_L^2 - p_R^2)/2 = \lambda \odot \lambda/2$ takes an integer value for any vector $\lambda$ in the even self-dual lattice $\Lambda_0$.
Also $\bar h=0$ imposes $\lambda_1 = \lambda_2$, so the primary operators with $\bar h=0$ are associated with the lattice vectors in $\Lambda_2$ of the form: $\lambda + \delta$ with $\lambda = (v,v) \in \Lambda(\CC)$, where $v$ is a vector in the \emph{Euclidean} Construction A lattice $\Lambda_\text{E}(C)$ defined by
\begin{align}
    \Lambda_\text{E}(C) 
        =
            \left\{ \frac{c + p\,m}{\sqrt{p}}\, \bigg|\, c\in C, ~ m\in \BZ^{n} \right\}\ ,
\end{align}
which is odd self-dual for $C$ a self-dual code.
Note that $\lambda + \delta \in \Lambda_2$ implies that $\lambda \in \Lambda_1$ when $n \in 4\BZ$ and $\lambda \in \Lambda_0$ when $n \in 4\BZ +2$. 
The former is, however, impossible as $\lambda$ satisfies $\chi\odot \lambda = 2\sqrt{p}\,v\cdot \mathbf{1}_n = 2(c+p\,m)\cdot  \mathbf{1}_n =0~\text{mod}~2$.
Thus, we focus on the case with $n\in 4\BZ +2$.
The momentum vectors for the primary operators with $\bar h=0$ in the NS sector take the form, $(p_L, p_R) = \left( \sqrt{2}\,u, 0\right)$, $u\in S(\Lambda_\text{E}(C))$, where $S(\Lambda_\text{E}(C))$ is the \emph{shadow}~\cite{conway1990new} of the Euclidean lattice $\Lambda_\text{E}(C)$ defined by $S(\Lambda_\text{E}(C)) =
        \Lambda_\text{E}(C) + \frac{\sqrt{p}}{2}\,\textbf{1}_n$.
The shadow determines the spectrum of the primary operators with $\bar h=0$ in the NS sector as $h = u^2$ with $u \in S(\Lambda_\text{E}(C))$, and allows us to test if there exist operators of $h=3/2$ once a self-dual classical code $C$ is given.

Next, we turn to the constraint imposed by condition (ii).
Let us focus on the left-moving momenta $p_L$ of the vertex operators in the R sector.
For $n\in 4\BZ +2$, they are in 
$\widetilde{\Lambda}_1\cup\widetilde{\Lambda}_3$.
The left-moving momenta are $p_L = \left(c_1 + c_2 + p\,(m_1 + m_2)\right)/\sqrt{2p}$ for $p_L \in \tilde\Lambda_1$ and $p_L = \left(c_1 + c_2 + p\,(m_1 + m_2 +\mathbf{1}_n)\right)/\sqrt{2p}$ for $p_L \in \tilde\Lambda_3$, where $c_1,c_2\in C\,,\;m_1,m_2\in\BZ^n$ and $\mathbf{1}_n\cdot(c_1+c_2)\neq \mathbf{1}_n\cdot(m_1+m_2)$ mod 2.
In both cases, using the linearity of the classical code $C$, we can represent them as $p_L = (c + p\,m)/\sqrt{2p}$, where $c\in C, m\in\BZ^n$ and $\mathbf{1}_n\cdot(c + m)=1$ mod 2.
Since $\mathbf{1}_n\cdot(c + m)  = \mathbf{1}_n\cdot(c + p\,m) = (c+p\,m)^2 = 1$ mod $2$ and $\left(c+p\,m\right)^2/p$ takes a non-negative integer value as $c\cdot c = 0$ mod $p$ for $c\in C$, the positive energy condition $h\ge \frac{n}{24}$ amounts to
\begin{align}
\label{eq:susy_cond_fi}
    \min_{\substack{v\in \Lambda_\text{E}(C)\\ ~v^2 = 1~\text{mod}\,2}} v^2 \geq \frac{n}{6}\ .
\end{align}
For a code of length $n\le 6$, \eqref{eq:susy_cond_fi} is always satisfied.
There are no further constraints from the positive energy condition for the right-moving sector $\bar h \ge \frac{n}{24}$ as it yields the same condition as \eqref{eq:susy_cond_fi} in the left-right symmetric theory.

To recapitulate the discussion so far, we have shown that a fermionic CFT built out of the CSS code associated with a self-dual classical code $C$ has a chance of possessing supersymmetry only if $C$ has length $n\in4\BZ +2$, a vector $u\in S(\Lambda_\text{E}(C))$ satisfies $u^2 = 3/2$, and the condition \eqref{eq:susy_cond_fi} is met.
Otherwise, the fermionic CFT cannot have supersymmetry.

Armed with this result, we are now in a position to search for supersymmetric CFTs by exploiting self-dual classical codes with the necessary properties.
Let us consider the cases with $n\le 6$ so that the condition \eqref{eq:susy_cond_fi} is automatically satisfied.
Since $n\in 4\BZ +2$, we have two cases; $n=2$ and $n=6$.
For $p=5$, there are one self-dual code $C_2$ of length $n=2$ and two self-dual codes $C_2^3$ and $F_6$ of length $n=6$ \cite{leon1982self}.
Let us examine the $C_2$ code which has five codewords $C_{2} = \{(0,0),(1,2),(2,4),(3,1),(4,3)\}$.
The Euclidean lattice $\Lambda_\text{E}(C_2)$ becomes a two-dimensional lattice isomorphic to $\BZ^2$, whose orthonormal basis can be chosen as $v_1 = (1,2)/\sqrt{5}$ and $v_2 = (2,-1)/\sqrt{5}$.
With this basis, we have $\frac{\sqrt{5}}{2}\,\textbf{1}_2 =  \left(3v_1 + v_2\right)/2$, and any vector $u$ in the shadow $S\left(\Lambda_\text{E}(C_{2})\right)$ can be written as $u = m_1'v_1 +m_2' v_2$ for $m_1'$, $m_2'\in \BZ+1/2$. 
In this case, there are no solutions for $u^2 = m_1'^2 + m_2'^2 = 3/2$ and thus the fermionic CFT cannot be supersymmetric.
Next, consider the code $C_{2}^3 = C_2 \times C_{2} \times C_{2}$.
The shadow is also decomposed into the direct product of three $S\left(\Lambda_\text{E}(C_{2})\right)$.
One can expand any vector $u\in \Lambda_\text{S}(C_2^3)$ as $u=\sum_{i=1}^6 m_i'\,r_i$ for $m_i' \in \BZ + 1/2~(i=1, \cdots, 6)$ in the orthonormal basis $r_i$ given by $r_1 = (v_1, 0^4), r_2 = (v_2, 0^4), r_3 = (0^2, v_1, 0^2), r_4 = (0^2, v_2, 0^2), r_5 = (0^4, v_1), r_6 = (0^4, v_2)$ where $0^{k} = (0,0,\cdots, 0)$ is the all-zero vector of length $k$.
Solving the equation $u^2 = \sum_{i=1}^6 m_i'^2 = 3/2$, we find $64$ solutions $m_i' = \pm 1/2~(i=1,\cdots, 6)$.
Thus this theory meets the both conditions $(\text{i}), (\text{ii})$.
We also checked numerically the condition (iii): $Z_\text{RR} = T - U = 24$.
Therefore, this model is potentially supersymmetric.

We now show that the fermionic CFT constructed from~$C_2^3$ indeed has supersymmetry.
By construction, the lattice $\Lambda(\mathcal{C})$ given in~(\ref{eq:Lambda-CSS}) has a basis given by $(r_i,0^6)$ and $(0^6,r_i)$ with $i=1,\cdots,6$, and is therefore isomorphic to 
$\mathbb{Z}^{12}$
with Lorentzian metric $\eta$ in~(\ref{eq:metric}).
Because the coefficients in the expansion $\chi=\sqrt{5}\mathbf{1}_{12}= \sum_{i=1}^3(3 r_{2i-1}+r_{2i},3 r_{2i-1}+r_{2i})$ 
are all odd, the rotation defined by the basis gives isomorphisms
\begin{align}
&    \Lambda_i \simeq  \Lambda_i^{(0)} = \bigg\{ m\in
\mathbb{Z}^{12}
 \bigg|  \sum_{j=1}^{12} m_j = i\;\;\mathrm{mod}\;2 \bigg\}
    \,,
    \\
&    
\qquad\qquad
\Lambda_{i+2} \simeq \Lambda_{i+2}^{(0)} = \Lambda_i^{(0)} + \frac{1}{2} \mathbf{1}_{12}\,,
\end{align}
for $i=0,1$.
Thus, after the rotation, the left- and right-moving momenta $(p_L,p_R)$ take values in
\begin{align}
&\hspace{-2mm}
\widetilde{\Lambda}^{(0)}_i 
\hspace{-0.5mm}
=
\hspace{-0.5mm}
\bigg\{ 
\hspace{-0.5mm}
\frac{(a,a)}{\sqrt{2}} + \sqrt{2}m \bigg| a\in
\mathbb{F}_2^6
\,, 
m\in
\mathbb{Z}^{12}
, \sum_{j=1}^6 a_j = i 
\bigg\}
,
\hspace{-0.5mm}
\\
&\qquad\qquad\qquad
\widetilde{\Lambda}^{(0)}_{i+2} 
=
\widetilde{\Lambda}^{(0)}_i + \frac{1}{\sqrt{2}}(\mathbf{1}_6,0^6) \,,
\end{align}
for $i=0,1$.
We observe that the (shifted) lattices $\widetilde{\Lambda}^{(0)}_i$ ($i=0,1,2,3$) are precisely the momentum lattices in the $\widehat{\mathfrak{su}}(2)_1^6$ description of the K3 sigma model~\cite{Gaberdiel:2013psa}, which we refer to as the GTVW model. 
In particular, the fermionic CFT has $\mathcal{N}=(4,4)$ supersymmetries whose currents were explicitly constructed in~\cite{Gaberdiel:2013psa} and  interpreted in terms of quantum error correcting codes in~\cite{Harvey:2020jvu}.

The other code $F_6$ is generated by  three codewords~\cite{leon1982self}
$(1,0,1,-1,-1,1)$, $(1^2,0,1,-1,-1)$,  $(1,-1,1,0,1,-1)$.
The lattice~$\Lambda(\mathcal{C})$ for $F_6$ has a basis given by $(s_i,0^6)$ and $(0^6,s_i)$ with $s_i=\tilde{s}_i/\sqrt{5}$, where $\tilde{s}_1=(0,1^5)$, $\tilde{s}_2=(1,0,1,-1,-1,1)$, $\tilde{s}_3=(1,-1,-1,1,0,1)$, $\tilde{s}_4=(1,-1,1,0,1,-1)$, $\tilde{s}_5=(1^2,0,1,-1,-1)$, and $\tilde{s}_6=(1^2,-1,-1,1,0)$.  
The odd coefficients in $\chi=(5s_1+s_2+\cdots+s_6,5s_1+s_2+\cdots+s_6)$ again imply that the fermionic CFT constructed from $F_6$ coincides with the GTVW model~\cite{Gaberdiel:2013psa}.
We note that, in general, different codes can yield the same CFT as exemplified by $C_2^3$ and~$F_6$.

\section{Discussion}

In this letter, we revealed novel relations
among fermionic CFTs, quantum stabilizer codes, and lattices together with their modifications.
We also examined the necessary conditions for the fermionic code CFTs to be supersymmetric and found two CSS codes defined by classical self-dual codes with $p=5$ and $n=6$ that satisfy the conditions.
We further proved that these CFTs are nothing but the 
GTVW model~\cite{Gaberdiel:2013psa}, which has $\mathcal{N}=4$ supersymmetry.
Given the successful application of our method, we expect that more supersymmetric CFTs can be constructed from quantum stabilizer codes.
Focusing on CSS codes defined by classical self-dual codes, fermionic code CFTs can be supersymmetric only when $n\in 4\BZ + 2$.
When $p=5$, there exist classical self-dual codes for even $n$ \cite{leon1982self}, and some of them may give rise to supersymmetric code CFTs if the necessary conditions given in the text are met.
For $n=10$, fermionic code CFTs cannot be supersymmetric as the associated Construction A lattices have lattice vectors of length one~\cite{conway1998note}, which violate the condition \eqref{eq:susy_cond_fi}.
On the other hand, there exist self-dual codes of length $n=14$ that meet the necessary conditions for supersymmetry, thus the associated code CFTs are candidates of supersymmetric CFTs.

We have focused on qudit stabilizer codes based on $\BF_p$ for an odd prime $p$. A similar discussion can be applied to qubit stabilizer codes ($p=2$)\,\footnote{We also expect that our construction of fermionic CFTs can be applied to a broader class of Narain code CFTs including those considered in \cite{Angelinos:2022umf}.}. In the binary case, the choice of $\chi\in\Lambda(\CC)$ depends on the type of classical code $\CC$. In what follows, we focus on the CSS construction $\CC = C\times C$ for a binary self-dual code $C$. Then we can take $\chi =2\delta= \mathbf{1}_{2n}/\sqrt{2}$ where the non-anomalous condition~\cite{Lin:2019kpn} imposes $n\in4\BZ$.
While the $\BZ_2$ grading of $\Lambda(\CC)$ is the same as in the previous case \eqref{eq:lambda0lambda1}, $\Lambda_2$ and $\Lambda_3$ are slightly modified:
\begin{align}
    (\Lambda_2, \Lambda_3) &= 
    \begin{dcases}
    (\Lambda_1 + \delta, \Lambda_0 + \delta) & \quad (n\in8\BZ)\,,\\
    (\Lambda_0 + \delta, \Lambda_1 + \delta) & \quad (n\in 8\BZ+4)\,.
    \end{dcases}
\end{align}
The sectors after fermionization take the same form as in Table \ref{tab:fermionization}.
Our construction provides examples of fermionic CFTs satisfying the conditions (i)-(iii) for supersymmetry.
One example is given by the unique self-dual code $B_4$ ($C_2^2$ of~\cite{pless1972classification}) of length $4$ generated by two 
codewords
$(1^2, 0^2)$ and $(0^2, 1^2)$,
and another by the unique indecomposable self-dual code $B_{12}$~\cite{pless1972classification} of length 12 generated by six 
codewords
$(0^2, 1,0,1,0^2,1,0^3,1)$, $(1,0^3,1,0^3,1^4)$, $(0^6,1^2,0^2,1^2)$, $(0,1,0^2,1,0^2,1^4,0)$, $(0^3, 1,0^3,1^2,0^2,1)$, and $(0^5,1,0,1,0,1,0,1)$.
For the self-dual code $B_4$, the fermionic code CFT contains 16 Virasoro primaries of weights $(3/2,0)$
and the RR partition function vanishes.
On the other hand, the fermionic code CFT from $B_{12}$ contains 64 Virasoro primaries of weights $(3/2,0)$,
and the resulting RR partition function takes the constant value $288$. 
These observations strongly suggest the existence of supersymmetry in both cases. 
It remains open whether they are equivalent to known models with supersymmetry or provide new examples of supersymmetric CFTs.

In general, it is nontrivial to confirm the existence of supersymmetry in a given candidate CFT as it requires the explicit construction of supercurrents as a linear combination of vertex operators of weight $3/2$.
In the GTVW model, a supercurrent operator that generates a part of the supersymmetry can be represented as a linear combination of $2^6$ primary operators. This structure has been identified with a  $[[6,0]]$ qubit stabilizer code \cite{Harvey:2020jvu}, where the supercurrent is viewed as a one-dimensional code subspace in the space of 6 qubits. 
It is desirable to
find a connection, if any,
between their interpretation and our construction of the GTVW model.

For bosonic Narain code CFTs constructed from a class of CSS codes, we can exactly compute the averaged partition function~\cite{Kawabata:2022jxt}.
Our construction of the fermionized code CFTs enables us to take their average similarly.
These averaged theories may have a holographic description related to an abelian Chern-Simons theory both in bosonic~\cite{Maloney:2020nni,Afkhami-Jeddi:2020ezh} and fermionic~\cite{Ashwinkumar:2021kav} cases.
It would be interesting to give a holographic interpretation for our class of fermionic CFTs.

\bigskip
\begin{acknowledgments}
%\paragraph*{Acknowledgments.}
We are grateful to S.\,Yahagi for valuable discussions.
The work of T.\,N. was supported in part by the JSPS Grant-in-Aid for Scientific Research (C) No.19K03863, Grant-in-Aid for Scientific Research (A) No.\,21H04469, and
Grant-in-Aid for Transformative Research Areas (A) ``Extreme Universe''
No.\,21H05182 and No.\,21H05190.
The research of T.\,O. was supported in part by Grant-in-Aid for Transformative Research Areas (A) ``Extreme Universe'' No.\,21H05190. 
The work of K.\,K. was supported by FoPM, WINGS Program, the University of Tokyo.
\end{acknowledgments}

\bibliography{QEC_CFT}

\end{document}